\documentclass[preprint,nofootinbib,floatfix,amsmath,amssymb,showpacs,superscriptaddress]{revtex4}

\usepackage{graphicx}
\usepackage{dcolumn}
\usepackage{bm}
\usepackage{color}
\usepackage[bordercolor=white,backgroundcolor=gray!30,linecolor=black,colorinlistoftodos]{todonotes}

\newcommand{\xcc}{{\Xi_{cc}}}

\newcommand{\xrho}{\Xi_{cc}\Xi_{cc}\rho}
\newcommand{\ffac}[2]{F_{#1,{\cal #2}}}

\begin{document}

\title{Electromagnetic properties of doubly charmed baryons \\ in Lattice QCD}

\author{K. U. Can}
\author{G. Erkol}
\author{B. Isildak}
\affiliation{Department of Natural and Mathematical Sciences, Faculty of Engineering, Ozyegin University, Nisantepe Mah. Orman Sok. No:13, Alemdag 34794 Cekmekoy, Istanbul Turkey}
\author{M. Oka}%
\affiliation{Department of Physics, H-27, Tokyo Institute of Technology, Meguro, Tokyo 152-8551 Japan}
\author{T. T. Takahashi}
\affiliation{Gunma National College of Technology, Maebashi, Gunma 371‐8530, Japan }

\date{\today}

\begin{abstract}
We compute the electromagnetic properties of $\xcc$ baryons in 2+1 flavor Lattice QCD. By measuring the electric charge and magnetic form factors of $\xcc$ baryons, we extract the magnetic moments, charge and magnetic radii as well as the $\xcc\xcc\rho$ coupling constant, which provide important information to understand the size, shape and couplings of the doubly charmed baryons. We find that the two heavy charm quarks drive the  charge radii and the magnetic moment of $\xcc$ to smaller values as compared to those of, \emph{e.g.}, the proton.

\end{abstract}
\maketitle

\section{Introduction}

There has been recently a profound interest in charmed baryons which was mainly triggered by the experimental discovery of the doubly charmed baryons $\xcc$ by SELEX~\cite{Mattson:2002vu, Ocherashvili:2004hi}, although it was not confirmed by BABAR~\cite{Aubert:2006qw} and BELLE experiments~\cite{Chistov:2006zj}. While there are many states yet to be confirmed and discovered experimentally, the charmed baryon sector is under an intense theoretical investigation. One of the several aspects which makes the physics of doubly charmed baryons interesting is that the binding of two heavy quarks and a light quark provides a unique perspective for dynamics of confinement. Moreover, the weak decays of doubly charmed baryons give an insight to the dynamics of singly charmed baryons.

Previous calculations on the charmed baryons have been mostly concentrated on their spectrum. To this end, various methods have been used: quark models~\cite{Roberts:2007ni, Martynenko:2007je}, Heavy Quark Effective Theory~\cite{Korner:1994nh}, QCD Sum Rules~\cite{Groote:1996em, Zhang:2008pm}, Lattice QCD with quenched approximation~\cite{Flynn:2003vz, Mathur:2002ce, Lewis:2001iz} and with dynamical quarks~\cite{Liu:2009jc, Namekawa:2011wt, Alexandrou:2012xk, Briceno:2012wt, Namekawa:2013vu}. Recent lattice results predict the mass of the $\xcc$ baryon to be $\sim$100~MeV larger as compared to the experimental value by SELEX which is 3518.9(9)~MeV.

While the baryon spectrum still stands as a challenge for the lattice QCD simulations, the electromagnetic properties of baryons are as crucial, for deciphering the internal structure. Lattice computations can now probe the electromagnetic structure of the nucleon for pion masses as low as~$m_\pi\sim 180$~MeV~\cite{Collins:2011mk} and with a technology being continuously improved~\cite{Alexandrou:2011db}. $\xcc$ baryons are particularly interesting in this respect. Determining how the three quarks distribute themselves inside the baryon \-- when two of them are heavy \-- enhances our understanding of the heavy-quark dynamics and sheds light on the internal structures of, not only heavy, but all baryons.  

Our aim in this work is to compute the electromagnetic form factors, as well as the charge radii and the magnetic moments of the spin-1/2 $\xcc$ baryons in 2+1 flavor Lattice QCD. We also compute the $\xrho$ coupling constant, which is an important ingredient for the parameterization of charmed-baryon molecular states in terms of one-boson exchange potential models~\cite{Liu:2011xc, Meguro:2011nr}. Note that the magnetic moments of charmed baryons have been considered before in quark models~\cite{JuliaDiaz:2004vh,Faessler:2006ft,Albertus:2006ya} and QCD Sum Rules~\cite{Zhu:1997as, Aliev:2001ig}. Our work is organized as follows: In Section~\ref{sec2} we present the theoretical formalism of our calculations of the form factors together with the lattice techniques we have employed to extract them. In Section~\ref{sec3} we present and discuss our numerical results. Section~\ref{sec4} contains a summary of our findings.

\section{The formulation and the lattice simulations}\label{sec2}

To compute the electromagnetic form factors, we consider the baryon matrix elements of the electromagnetic vector current, $V_\mu=\frac{2}{3}\overline{c}\gamma_\mu c+\frac{2}{3}\overline{u}\gamma_\mu u-\frac{1}{3}\overline{d}\gamma_\mu d$, which can be written in the form
	\begin{equation}\label{matel}
	\langle {\cal B}(p)|V_\mu|{\cal B}(p^\prime)\rangle= \bar{u}(p) \left[\gamma_\mu F_{1,{\cal B}}(q^2) +i \frac{\sigma_{\mu\nu} q^\nu}{2m_{\cal B}} F_{2,{\cal B}}(q^2) \right]u(p),
	\end{equation}
where $q_\mu=p_\mu^\prime-p_\mu$ is the transferred four\--momentum. Here $u(p)$ denotes the Dirac spinor for the baryon with four-momentum $p^\mu$ and mass $m_{\cal B}$. The Sachs form factors $F_{1,{\cal B}}(q^2)$ and $F_{2,{\cal B}}(q^2)$ are related to the electric and magnetic form factors by
\begin{align}
	G_{E,\cal{B}}(q^2)=\ffac{1}{B}(q^2)+\frac{q^2}{4m_{\cal B}^2}\ffac{2}{B}(q^2),\\
	G_{M,\cal{B}}(q^2)=\ffac{1}{B}(q^2)+\ffac{2}{B}(q^2).
\end{align}

Our method of computing the matrix element in Eq.~\eqref{matel}, which was employed to extract the nucleon electromagnetic form factor, follows closely that of Ref.\cite{Alexandrou:2011db}. Using the following ratio
\begin{align}
\begin{split}\label{ratio}
	&R(t_2,t_1;{\bf p}^\prime,{\bf p};\Gamma;\mu)=\\
	&\quad\frac{\langle F^{{\cal B} {\cal V}_\mu {\cal B}}(t_2,t_1; {\bf p}^\prime, {\bf p};\Gamma)\rangle}{\langle F^{{\cal B}}(t_2; {\bf p}^\prime;\Gamma_4)\rangle} \left[\frac{\langle F^{{\cal B}}(t_2-t_1; {\bf p};\Gamma_4)\rangle}{\langle F^{{\cal B}}(t_2-t_1; {\bf p}^\prime;\Gamma_4)\rangle}\right.\\ 
	&\quad\left.\times\frac{\langle F^{{\cal B}}(t_1; {\bf p}^\prime;\Gamma_4)\rangle \langle F^{{\cal B}}(t_2; {\bf p}^\prime;\Gamma_4)\rangle}{\langle F^{{\cal B}}(t_1; {\bf p};\Gamma_4)\rangle \langle F^{{\cal B}}(t_2; {\bf p};\Gamma_4)\rangle} \right]^{1/2},
\end{split}
\end{align}
where the baryonic two-point and three-point correlation functions are respectively defined as:
\allowdisplaybreaks{
\begin{align}
	\begin{split}\label{twopcf}
	&\langle F^{{\cal B}}(t; {\bf p};\Gamma_4)\rangle=\sum_{\bf x}e^{-i{\bf p}\cdot {\bf x}}\Gamma_4^{\alpha\alpha^\prime} \\
	&\qquad\times \langle \text{vac} | T [\eta_{\cal B}^\alpha(x) \bar{\eta}_{{\cal B}}^{\alpha^\prime}(0)] | \text{vac}\rangle,
	\end{split}\\
	\begin{split}
	&\langle F^{{\cal B V_\mu B^\prime}}(t_2,t_1; {\bf p}^\prime, {\bf p};\Gamma)\rangle=-i\sum_{{\bf x_2},{\bf x_1}} e^{-i{\bf p}\cdot {\bf x_2}} e^{i{\bf q}\cdot {\bf x_1}} \\
	&\qquad\times \Gamma^{\alpha\alpha^\prime} \langle \text{vac} | T [\eta_{\cal B}^\alpha(x_2) V_\mu(x_1) \bar{\eta}_{{\cal B}^\prime}^{\alpha^\prime}(0)] | \text{vac}\rangle,
	\end{split}
\end{align}
}%
with $\Gamma_i=\gamma_i\gamma_5\Gamma_4$ and $\Gamma_4\equiv (1+\gamma_4)/2$. The $\xcc$ interpolating fields are chosen, similarly to that of nucleon, as
\begin{equation}
		\eta_{\xcc}(x)=\epsilon^{ijk}[c^{T i}(x) C \gamma_5 \ell^j(x)]c^k(x),
\end{equation}
where $\ell=u$ for the doubly charged $\Xi_{cc}^{++}(ccu)$ and $\ell=d$ for the singly charged $\Xi_{cc}^{+}(ccd)$ baryon. Here $i$, $j$, $k$ denote the color indices and $C=\gamma_4\gamma_2$. $t_1$ is the time when the external electromagnetic field interacts with a quark and $t_2$ is the time when the final baryon state is annihilated. When $t_2-t_1$ and $t_1\gg a$, the ratio in Eq.~(\ref{ratio}) reduces to the desired form
\begin{equation}\label{desratio}
	R(t_2,t_1;{\bf p^\prime},{\bf p};\Gamma;\mu)\xrightarrow[t_2-t_1\gg a]{t_1\gg a} \Pi({\bf p^\prime},{\bf p};\Gamma;\mu).
\end{equation}
We extract the form factors $G_{E,\cal{B}}(q^2)$ and $G_{M,\cal{B}}(q^2)$ by choosing appropriate combinations of Lorentz direction $\mu$ and projection matrices $\Gamma$:
\begin{align}
	\Pi({\bf 0},{\bf -q};\Gamma_4;\mu=4)&=\left[\frac{(E_{\cal B}+m_{\cal B})}{2E_{\cal B} }\right]^{1/2} G_{E,\cal{B}}(q^2),\\
	\Pi({\bf 0},{\bf -q};\Gamma_j;\mu=i)&=\left[\frac{1}{2E_{\cal B} (E_{\cal B}+m_{\cal B})}\right]^{1/2} \epsilon_{ijk}\, q_k\, G_{M,\cal{B}}(q^2).
\end{align}
Here, $G_{E,\cal{B}}(0)$ gives the electric charge of the baryon. Similarly, the magnetic moment can be obtained from the magnetic form factor $G_{M,\cal{B}}$ at zero momentum transfer.

Our lattice setup is explained in detail in Ref.~\cite{Can:2012tx}. We use the \emph{wall method} which does not require to fix sink operators in advance and hence allowing us to compute all baryon channels we are interested in simultaneously. However, since the wall sink/source is a gauge-dependent object, we have to fix the gauge, which we choose to be Coulomb. We extract the baryon masses from the two-point correlator with shell source and point sink, and use the dispersion relation to calculate the energy at each momentum transfer. 

Our simulations have been run on $32^3\times 64$ lattices with 2+1 flavors of dynamical quarks and we use the gauge configurations that have been generated by the PACS-CS collaboration~\cite{Aoki:2008sm} with the nonperturbatively $\mathcal{O}(a)$-improved Wilson quark action and the Iwasaki gauge action. We use the gauge configurations at $\beta=1.90$ with the clover coefficient $c_{SW}=1.715$ and they have a lattice spacing of $a=0.0907(13)$ fm ($a^{-1}=2.176(31)$~GeV). We consider four different hopping parameters for the sea and the $u$,$d$ valence quarks, $\kappa_{sea},\kappa_{val}^{u,d}=$ 0.13700, 0.13727, 0.13754 and 0.13770, which correspond to pion masses of $\sim$ 700, 570, 410, and 300~MeV. The hopping parameter for the $s$ sea quark is fixed to $\kappa_{sea}^{s}=0.1364$.

Similar to our simulations in Ref.~\cite{Can:2012tx}, we choose to employ Clover action for the charm quark. While this choice may seem questionable since the Clover action is subject to discretization errors of $\mathcal{O}(m_q\,a)$, the calculations which are insensitive to a change of charm-quark mass are less severely affected by these errors~\cite{Bali:2011rd, Can:2012tx}. In our case, we have estimated the effect of discretization errors to be of the order of a few percent (see the discussion below). Precision calculations such as the spectral properties and the hyperfine splittings requires a more careful treatment of these lattice artefacts by considering improved actions such as Fermilab~\cite{ElKhadra:1996mp}. Note that the Clover action we are employing here is a special case of the Fermilab heavy-quark action with $c_{SW}=c_E=c_B$~\cite{Burch:2009az}. We determine the hopping parameter of the charm quark ($\kappa_{c}=0.1224$) so as to reproduce the mass of J/$\psi(3097)$. 

We employ smeared source and wall sink which are separated by 12 lattice units in the temporal direction. Source operators are smeared in a gauge-invariant manner with the root mean square radius of $\sim 0.5$ fm. All the statistical errors are estimated via the jackknife analysis. In this work, we consider only the connected diagrams. Computation of the disconnected diagrams is a numerically demanding task. Their contributions, on the other hand, have been found to be consistent with zero in the case of nucleon electric form factors~\cite{Alexandrou:2012zz}. 

We make our measurements on 100, 100, 150 and 170 configurations, respectively for each quark mass. In order to increase the statistics we take several different source points using the translational invariance along the temporal direction. We make nine momentum insertions: $(|p_x|,|p_y|,|p_z|)=(0,0,0),(1,0,0),(1,1,0),(1,1,1),(2,0,0),(2,1,0),(2,1,1),(2,2,0),(2,2,1)$ and average over equivalent (positive and negative) momenta. In the case of magnetic form factors, we average over all possible equivalent combinations of momentum, spin projection and Lorentz component in order to increase the statistics. As a result the statistical precision is highly improved. We consider both the local 
\begin{equation}
	V_\mu=\overline{q}(x)\gamma_\mu q(x),
\end{equation}
and the point-split lattice vector current,
\begin{equation}
V_\mu = 1/2[\overline{q}(x+\mu)U^\dagger_\mu(1+\gamma_\mu)q(x) -\overline{q}(x)U_\mu(1-\gamma_\mu)q(x+\mu)],
\end{equation}
which is conserved by Wilson fermions, hence does not require any renormalization on the lattice. Since we obtain completely consistent results for both currents we present our results only for the point-split one.

\section{Results and Discussion}\label{sec3}

\begin{figure}[th]
	\centering
	\includegraphics[width=0.9\textwidth]{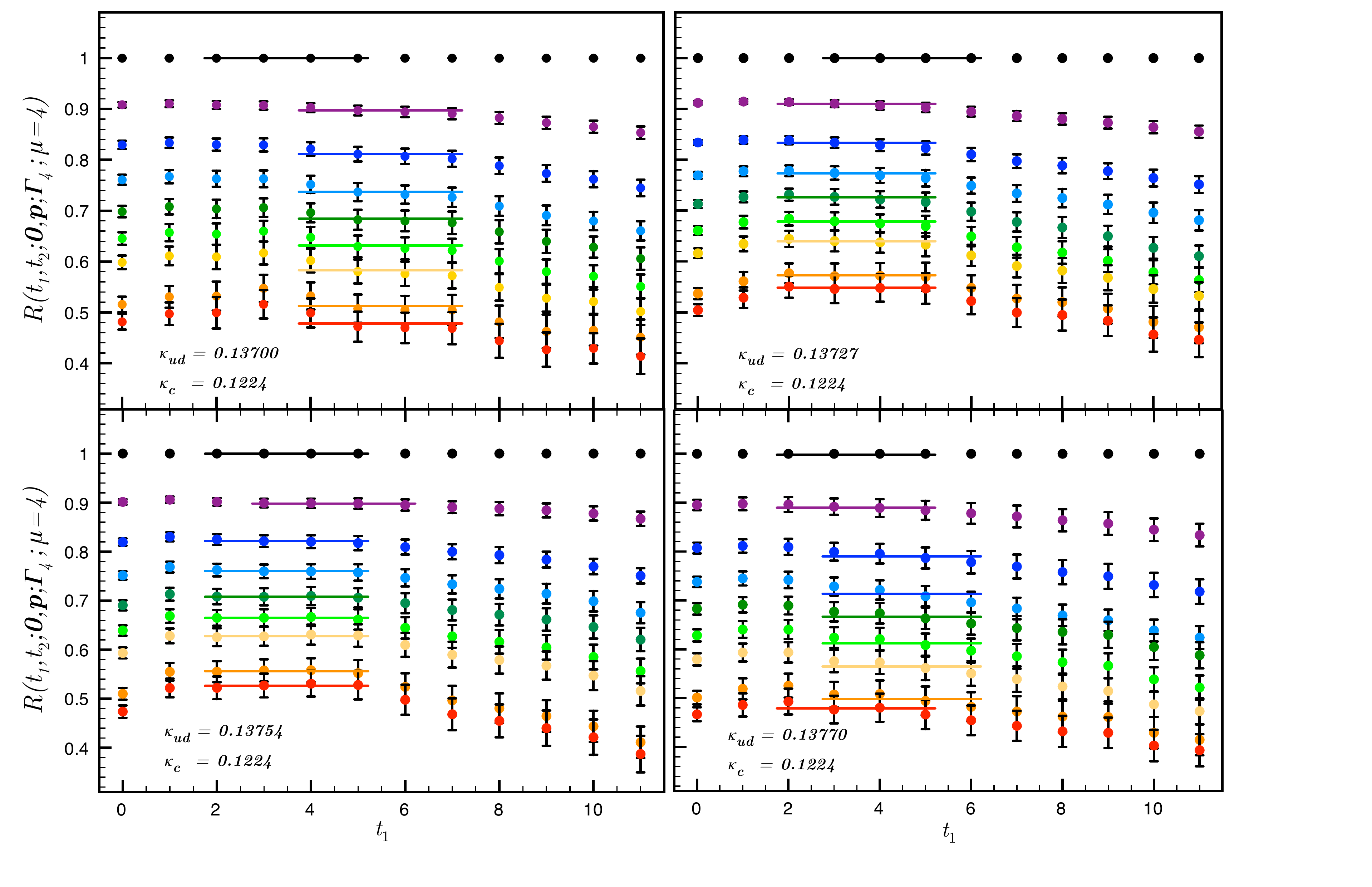}
	\caption{\label{el_plat} The ratio in Eq.~\eqref{ratio} as function of the current insertion time, $t_1$, for all the quark masses we consider and first nine four\--momentum insertions. The horizontal lines denote the plateau regions as determined by using a p-value criterion (see text).}
\end{figure}	

\begin{figure}[th]
	\centering
	\includegraphics[width=0.9\textwidth]{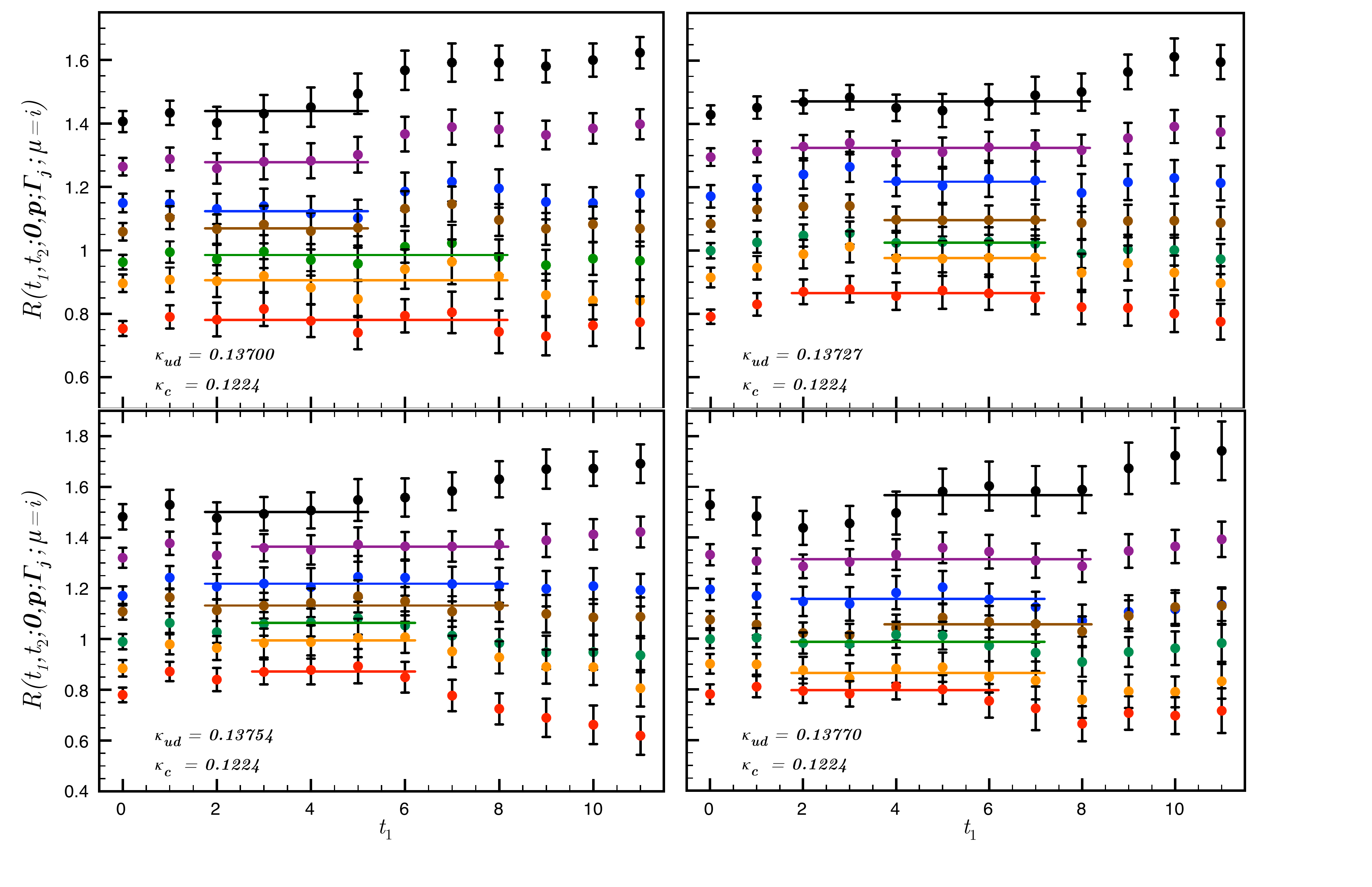}
	\caption{\label{mag_plat} Same as Fig.~\ref{el_plat} but for the magnetic form factors.}
\end{figure}	

We give the ratio in Eq.~\eqref{ratio} for the electric (magnetic) form factors as functions of the current insertion time, $t_1$, for each quark-mass value we consider and for the first nine (seven) momentum insertions in Fig.~\ref{el_plat} (Fig.\ref{mag_plat}). In determining a plateau region, we consider the p-value as a criterion.~\footnote{The p-value is the probability of having a $\chi^2$ value greater than or equal to the that obtained in the fit. Therefore a large p-value	indicates a stronger compatibility between the data and the fit form~\cite{PhysRevD.86.010001}.} In each case, we search for plateau regions of minimum three time slices between the source and the sink, and we choose the one that has the highest p-value. The regions closer to the smeared source are preferred as they are expected to couple to the ground state with higher strength as compared to the wall sink.

\begin{figure}[th]
	\centering
	\includegraphics[width=0.6\textwidth]{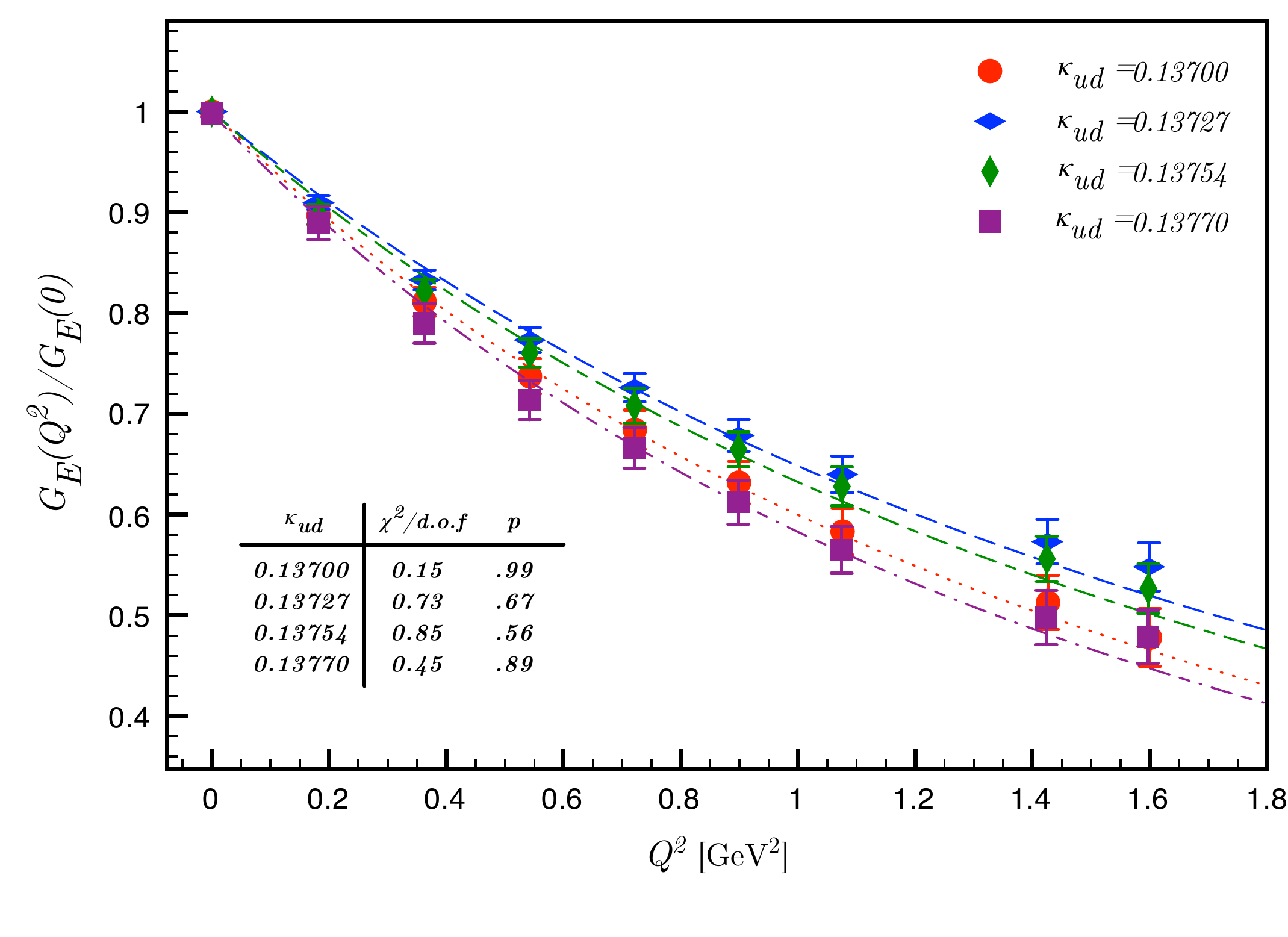}
	\caption{\label{eff_dipol} The electric form factor $G_E(Q^2)$ of $\Xi^{++}_{cc}$ as a function of $Q^2$ and as normalized with its electric charge, for all the quark masses we consider. The dots mark the lattice data and the curves show the best fit to the dipole form in Eq.~\eqref{dipole}.}
\end{figure}	

\begin{figure}[th]
	\centering
	\includegraphics[width=0.6\textwidth]{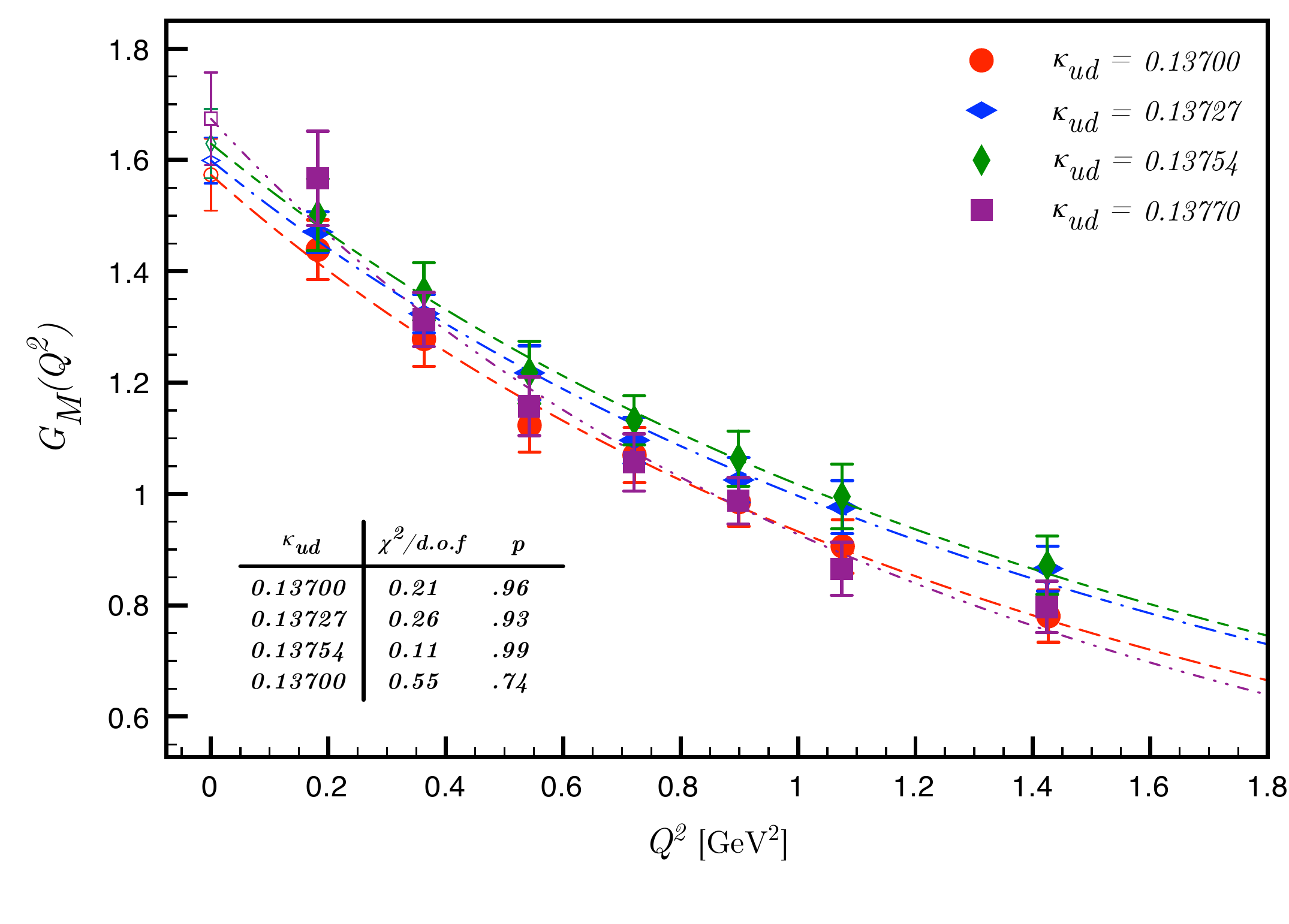}
	\caption{\label{mff_dipol} The magnetic form factor $G_M(Q^2)$ of $\Xi^{+}_{cc}$ as a function of $Q^2$ for all the quark masses we consider. The dots mark the lattice data and the curves show the best fit to the dipole form in Eq.~\eqref{dipole}.}
\end{figure}	

\begin{figure}[th]
	\centering
	\includegraphics[width=0.6\textwidth]{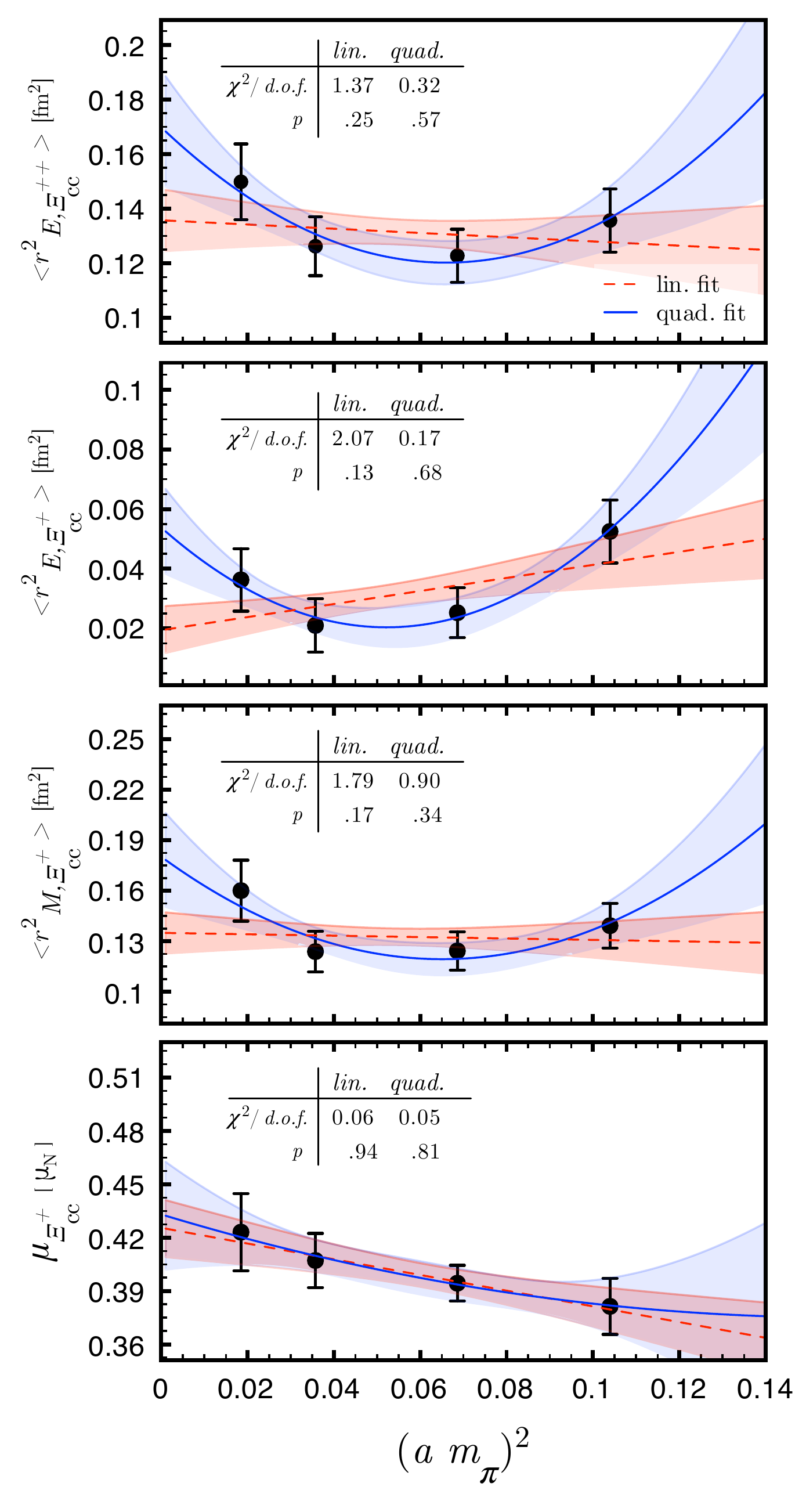}
	\caption{\label{chiral_fit} The chiral extrapolations for electric charge radius, magnetic charge radius and magnetic moment of $\xcc$ baryons in $(am_\pi)^2$. We show the fits to linear and quadratic forms. The straight and the dashed lines give the best fit and the shaded regions are the maximally allowed error regions. We also report the $\chi^2$ per degree of freedom and p-values for each fit.}
\end{figure}	

In order to obtain the magnetic moment, we need to extract the magnetic form factor $G_M$ at $-q^2 \equiv Q^2=0$. In other words, the lattice results need to be extrapolated to $Q^2=0$. We use a dipole form to describe the data at finite momentum transfers and to extrapolate:
\begin{equation}\label{dipole}
	G_{E,M}(Q^2)=\frac{G_{E,M}(0)}{(1+Q^2/\Lambda_{E,M}^2)^2}.
\end{equation}
Note that $G_E(0)=2$ for $\Xi^{++}_{cc}$ and $G_E(0)=1$ for $\Xi^{+}_{cc}$, which are obtained in our simulations to a very good accuracy. In Fig.~\ref{eff_dipol} (Fig.~\ref{mff_dipol}), we give the electric (magnetic) form factor of $\Xi^{++}_{cc}$ ($\Xi^{+}_{cc}$) as a function of $Q^2$. We show the lattice data and the fitted dipole forms for all the quark masses we consider. As can be seen from the figures, the dipole form describes the lattice data quite successfully with high-quality fits.

We can evaluate the electromagnetic charge radius of the $\xcc$ baryons from the slope of the form factor at $Q^2=0$,
\begin{equation}
	\langle r_{E,M}^2 \rangle=-\frac{6}{G_\text{E,M}(0)} \frac{d}{dQ^2}G_{E,M}(Q^2)\bigg|_{Q^2=0}.
\end{equation}
For the dipole form in Eq.~\eqref{dipole} we have
\begin{equation}\label{emfitform}
	\langle r_{E,M}^2 \rangle=\frac{12}{\Lambda_{E,M}^2},
\end{equation}
and the charge radii can be directly calculated using the values given in Table~\ref{res_table}.

To extract the coupling constant to the $\rho$-meson we use the VMD approach~\cite{Sakurai:69, Can:2012tx}:
\begin{equation}
	F_V(Q^2)=\frac{m_\rho^2}{m_\rho^2+Q^2}\frac{g_{\xcc\xcc\rho}}{g_\rho},
\end{equation}
where $m_\rho$ is the $\rho$-meson mass, $g_\rho$ is a constant which determines the coupling of the vector meson to the photon and we find $g_\rho=4.96$~\cite{Can:2012tx}.

The magnetic moment is defined as $\mu_{\cal B}=G_{M}(0) e/(2m_\xcc)$ in natural units. We obtain $G_{M}(0)$ by extrapolating the lattice data to $Q^2=0$ via the dipole form in Eq.~\eqref{dipole} as explained above. We evaluate the magnetic moments in nuclear magnetons using the relation
\begin{equation}
	\mu_\xcc=G_{M}(0)\left(\frac{e}{2m_\xcc}\right)=G_{M}(0)\left(\frac{m_N}{m_\xcc}\right)\mu_N,
\end{equation}
where $m_N$ is the physical nucleon mass and $m_\xcc$ is the $\xcc$ mass as obtained on the lattice.

\begin{table*}[ht]
	\caption{The dipole masses of electric and magnetic form factors $\Lambda_{E,M}$ in GeV, the electric and magnetic charge radii in fm$^2$, the values of magnetic form factors at $Q^2=0$ ($G_{M,\xcc}(0)$), the magnetic moment of $\Xi^{+}_{cc}$ ($\mu_{\Xi^{+}_{cc}}$) in nuclear magnetons,  $\xcc\xcc\rho$ coupling constant ($g_{\xcc\xcc\rho}$) and the mass of $\xcc$ in lattice units ($a\,m_{\xcc}$), at each quark mass we consider.
}
\begin{center}
\begin{tabular*}{1.0\textwidth}{@{\extracolsep{\fill}}cccccccc}
		\hline\hline 
		$\kappa^{u,d}_{val}$   &  $\Lambda_{E, \Xi^{++}_{cc}}$ & $\Lambda_{E, \Xi^{+}_{cc}}$ 
		 &   $\langle r_{E,\Xi^{++}_{cc}}^2 \rangle$ & $\langle r_{E,\Xi^{+}_{cc}}^2 \rangle$ & $a\,m_{\xcc}$    \\
		\hline \hline
		        & [GeV]     & [GeV]         & [fm$^2$]   & [fm$^2$]  &         \\
		0.13700 & 1.853(79) & 2.988(280)    & 0.136(12)  & 0.052(10) & 1.779(6) \\
		0.13727 & 2.033(71) & 4.183(644)    & 0.113(8)   & 0.027(8)  & 1.748(6) \\
		0.13754 & 1.970(75) & 4.723(955)    & 0.120(9)   & 0.021(8)  & 1.726(7) \\
		0.13770 & 1.801(81) & 3.555(500)    & 0.144(13)  & 0.037(10) & 1.706(8) \\
		\hline
		Lin. Fit & 1.697(63)& 3.600(306)    & 0.135(11)  & 0.020(7)    & 1.697(6)\\
		Quad. Fit& 1.476(103)& 2.409(698)   & 0.164(18)  & 0.049(12)   & 1.696(12) \\
		\hline\hline 
		$\kappa^{u,d}_{val}$ 
		& $\Lambda_{M, \Xi^{+}_{cc}}$ & $\langle r_{M,\Xi^{+}_{cc}}^2 \rangle$ & $G_{M,\Xi^{+}_{cc}}(0)$  &  $\mu_{\Xi^{+}_{cc}}$   & $g_{\xcc\xcc\rho}$    \\
		\hline \hline
		        & [GeV]      & [fm$^2$]  &           & [$\mu_N$] &  \\
		0.13700 & 1.829(86)  & 0.139(13) & 1.573(65) & 0.381(16) & 6.555(428)\\
		0.13727 & 1.936(88)  & 0.124(11) & 1.600(41) & 0.394(10) & 5.651(211)\\
		0.13754 & 1.940(94)  & 0.124(12) & 1.629(62) & 0.407(15) & 5.721(263) \\
		0.13770 & 1.706(97)  & 0.160(18) & 1.674(83) & 0.423(22) & 5.536(230) \\
		\hline
		Lin. Fit & 1.788(67) & 0.135(12) & 1.671(61) & 0.424(15) & 5.700(245)\\
		Quad. Fit &1.444(136)& 0.173(25) & 1.695(106)& 0.430(27) & 6.098(380) \\
		\hline\hline
\end{tabular*}
	\label{res_table}
\end{center}
\end{table*}

Our numerical results are given in Table~\ref{res_table}. We give the dipole masses of electric and magnetic form factors $\Lambda_{E,M}$, the electric and magnetic charge radii, the values of the magnetic form factors at $Q^2=0$, the magnetic moments of $\xcc$, $\xcc\xcc\rho$ coupling constant ($g_{\xcc\xcc\rho}$) and the mass of $\xcc$, at each quark-mass value we consider. In the case of the doubly charged $\Xi_{cc}^{++}$ baryon, the contributions of the $c$-quark and the $u$-quark to the magnetic moment are of the same size but opposite in sign. Such a cancellation of the quark contributions results in a much smaller value as compared to the magnetic moment of $\Xi_{cc}^{+}$, but then the data is somewhat noisy and inconclusive. Therefore we do not present the numerical results for the magnetic moment of $\Xi_{cc}^{++}$ here.

To obtain the values of the observables at the chiral point, we perform fits that are linear and quadratic in $m_{\pi}^2$:
\begin{align}
	&f_\text{lin}=a_1\,m_\pi^2+b_1,\\
	&f_\text{quad}=a_2\,m_\pi^4+b_2\,m_\pi^2+c,
\end{align}
where $a_{1,2},b_{1,2},c$ are the fit parameters. It is interesting to compare our result for the $\xcc$ mass with the one obtained by the PACS-CS from the same lattices; it must be noted that they use a relativistic heavy-quark action to keep the $\mathcal{O}(m_Q a)$ errors under control and find $a m_\xcc=1.656(10)$. A caveat is that PACS-CS extracts the $\xcc$ mass at the physical point without any chiral extrapolation, which in our case needs to be taken into account as a source of systematic error. Yet, a mass determination, of course, requires a more systematic chiral fit than linear or quadratic forms as we perform here. However, such a comparison is useful to see the effect of the discretization errors which are found to be of the order of a few percent. In addition to our findings in the meson sector we explicitly checked the sensitivity of the $\xcc$ form factors on charm-quark hopping parameter for the $\kappa_{ud}=0.13700$ lattices. We have found that changing the valence charm quark mass results in an approximately 100 MeV ($3\%)$ deviation in the $\xcc$ mass. Nevertheless this results in only a negligible change in the form factors. The electric charge radius deviates by less than $2\%$. We find $\langle r^2_{E,\xcc} \rangle_{\kappa_c = 0.1232} = 0.156(28)$ as compared to the $\langle r^2_{E,\xcc} \rangle_{\kappa_c = 0.1224} = 0.154(28)$ for 30 configurations. Considering also our final errors for the charge radius, it is safe to assume that these observables are insensitive to such a change in the charm quark mass at this precision.

Having improved the statistical precision by combinations of all possible momentum, spin projection and Lorentz components, the lattice data nicely extrapolate to the chiral point. In Fig.~\ref{chiral_fit}, we show the chiral extrapolations for the electric radius, charge radius and the magnetic moment of $\xcc$ baryons. The results of the two fit forms, namely linear and quadratic, with their error bands are given. In order to evaluate the quality of the fits, we report their $\chi^2$ per degree of freedom value and the p-values. While the fit results to linear and quadratic forms deviate from each other with their one to two standard deviations in some cases, the $\chi^2$ per degree of freedom and the p-values suggest that the quadratic form is favored over the linear form except for the magnetic moment.

We find that the electric charge radius of $\xcc$ is much smaller as compared to that of the proton (the experimental value is $\langle r_{E,p}^2 \rangle=$0.770~fm$^2$~\cite{PhysRevD.86.010001}) and this is in accordance with our conclusion in our recent work on D mesons~\cite{Can:2012tx} that the large mass of the $c$ quark drives the charge radii of charmed hadrons to smaller values. In Ref.~\cite{Brodsky:2011zs}, based on the SELEX doubly heavy baryon data, the authors show that the large isospin splitting implies a smaller electric charge radius, hence more compact baryon. It is relevant to note that our results for the charge radii are consistent with their findings. 

A similar conclusion can be reached for the magnetic properties of the charmed baryons. The magnetic moment and the magnetic charge radius of $\Xi^{+}_{cc}$ are much smaller as compared to those of the proton, which are $\mu_p=2.793~\mu_N$ and $\langle r_{M,p}^2\rangle=0.604$~fm$^2$~\cite{PhysRevD.86.010001}, respectively. As compared to the relativistic and nonrelativistic quark models, which predict $\mu_{\Xi^{+}_{cc}}\simeq 0.7\text{-}0.8~\mu_N$~\cite{JuliaDiaz:2004vh, Faessler:2006ft,Albertus:2006ya}, the computed value on the lattice is smaller. On the other hand, it is interesting to note that the charge radii do not systematically decrease as the pion mass increases, in opposite to what has been found for the nucleon. While this seems to contradict with our conclusion that the charge radii are smaller for heavier quarks, such behavior may be related to the modification of the confinement force in hadrons: The two charm quarks are compact in the $\xcc$ and the effect of the extra light quark is to modify the string tension between the two-charm component~\cite{Yamamoto:2007nn}. Further investigation is necessary to have a vigorous conclusion since this effect may change the inner dynamics of the hadron as the light quark gets heavier. Along this line, a lattice investigation of the doubly charmed and strange $\Omega_{cc}$ baryon seems timely as it can shed light on this issue.

\section{Conclusion}~\label{sec4}

We have computed the electromagnetic properties of the doubly charmed $\xcc$ baryons in 2+1-flavor lattice QCD for the first time in the literature. In particular, we have extracted the electric and magnetic charge radii and the magnetic dipole moment of $\xcc$ baryons, which provide very useful information about the size and the shape of the baryons. We have computed the form factors up to $\sim$1.5~GeV$^2$ and from these we have extracted the static electromagnetic properties. We have found that the two heavy quarks drive the electric and magnetic charge radii of the $\xcc$ baryon to lower values, as compared to, \emph{e.g.} the proton. In addition, we extracted the $\xrho$ coupling constant, which plays an important role in the description of charmed-baryon molecular states in terms of one-boson exchange potential models. Having explored the sector of doubly charmed baryons on the lattice, an interesting comparison can be made with the static electromagnetic properties of the singly charmed baryons, which would provide an informative perspective on the QCD dynamics at the heavy-quark level. A work along this line is still in progress.
 
\acknowledgments
G.~E. thanks Haris Panagopoulos and K.~U.~C. thanks C. McNeile, C. H\"{o}lbling and S. D\"{u}rr for invaluable comments and discussions. The numerical calculations in this work were performed on National Center for High Performance Computing of Turkey (Istanbul Technical University) under project number 10462009. The unquenched gauge configurations employed in our analysis were generated by PACS-CS collaboration~\cite{Aoki:2008sm}. We used a modified version of Chroma software system~\cite{Edwards:2004sx}. This work is supported in part by The Scientiﬁc and Technological Research Council of Turkey (TUBITAK) under project number 110T245 and in part by KAKENHI under Contract Nos. 22105503, 24540294 and 22105508.


\begin{thebibliography}{36}
\expandafter\ifx\csname natexlab\endcsname\relax\def\natexlab#1{#1}\fi
\expandafter\ifx\csname bibnamefont\endcsname\relax
  \def\bibnamefont#1{#1}\fi
\expandafter\ifx\csname bibfnamefont\endcsname\relax
  \def\bibfnamefont#1{#1}\fi
\expandafter\ifx\csname citenamefont\endcsname\relax
  \def\citenamefont#1{#1}\fi
\expandafter\ifx\csname url\endcsname\relax
  \def\url#1{\texttt{#1}}\fi
\expandafter\ifx\csname urlprefix\endcsname\relax\def\urlprefix{URL }\fi
\providecommand{\bibinfo}[2]{#2}
\providecommand{\eprint}[2][]{\url{#2}}

\bibitem[{\citenamefont{Mattson et~al.}(2002)}]{Mattson:2002vu}
\bibinfo{author}{\bibfnamefont{M.}~\bibnamefont{Mattson}} \bibnamefont{et~al.}
  (\bibinfo{collaboration}{SELEX Collaboration}),
  \bibinfo{journal}{Phys.Rev.Lett.} \textbf{\bibinfo{volume}{89}},
  \bibinfo{pages}{112001} (\bibinfo{year}{2002}), \eprint{hep-ex/0208014}.

\bibitem[{\citenamefont{Ocherashvili et~al.}(2005)}]{Ocherashvili:2004hi}
\bibinfo{author}{\bibfnamefont{A.}~\bibnamefont{Ocherashvili}}
  \bibnamefont{et~al.} (\bibinfo{collaboration}{SELEX Collaboration}),
  \bibinfo{journal}{Phys.Lett.} \textbf{\bibinfo{volume}{B628}},
  \bibinfo{pages}{18} (\bibinfo{year}{2005}), \eprint{hep-ex/0406033}.

\bibitem[{\citenamefont{Aubert et~al.}(2006)}]{Aubert:2006qw}
\bibinfo{author}{\bibfnamefont{B.}~\bibnamefont{Aubert}} \bibnamefont{et~al.}
  (\bibinfo{collaboration}{BABAR Collaboration}), \bibinfo{journal}{Phys.Rev.}
  \textbf{\bibinfo{volume}{D74}}, \bibinfo{pages}{011103}
  (\bibinfo{year}{2006}), \eprint{hep-ex/0605075}.

\bibitem[{\citenamefont{Chistov et~al.}(2006)}]{Chistov:2006zj}
\bibinfo{author}{\bibfnamefont{R.}~\bibnamefont{Chistov}} \bibnamefont{et~al.}
  (\bibinfo{collaboration}{BELLE Collaboration}),
  \bibinfo{journal}{Phys.Rev.Lett.} \textbf{\bibinfo{volume}{97}},
  \bibinfo{pages}{162001} (\bibinfo{year}{2006}), \eprint{hep-ex/0606051}.

\bibitem[{\citenamefont{Roberts and Pervin}(2008)}]{Roberts:2007ni}
\bibinfo{author}{\bibfnamefont{W.}~\bibnamefont{Roberts}} \bibnamefont{and}
  \bibinfo{author}{\bibfnamefont{M.}~\bibnamefont{Pervin}},
  \bibinfo{journal}{Int.J.Mod.Phys.} \textbf{\bibinfo{volume}{A23}},
  \bibinfo{pages}{2817} (\bibinfo{year}{2008}), \eprint{0711.2492}.

\bibitem[{\citenamefont{Martynenko}(2008)}]{Martynenko:2007je}
\bibinfo{author}{\bibfnamefont{A.}~\bibnamefont{Martynenko}},
  \bibinfo{journal}{Phys.Lett.} \textbf{\bibinfo{volume}{B663}},
  \bibinfo{pages}{317} (\bibinfo{year}{2008}), \eprint{0708.2033}.

\bibitem[{\citenamefont{Korner et~al.}(1994)\citenamefont{Korner, Kramer, and
  Pirjol}}]{Korner:1994nh}
\bibinfo{author}{\bibfnamefont{J.}~\bibnamefont{Korner}},
  \bibinfo{author}{\bibfnamefont{M.}~\bibnamefont{Kramer}}, \bibnamefont{and}
  \bibinfo{author}{\bibfnamefont{D.}~\bibnamefont{Pirjol}},
  \bibinfo{journal}{Prog.Part.Nucl.Phys.} \textbf{\bibinfo{volume}{33}},
  \bibinfo{pages}{787} (\bibinfo{year}{1994}), \eprint{hep-ph/9406359}.

\bibitem[{\citenamefont{Groote et~al.}(1997)\citenamefont{Groote, Korner, and
  Yakovlev}}]{Groote:1996em}
\bibinfo{author}{\bibfnamefont{S.}~\bibnamefont{Groote}},
  \bibinfo{author}{\bibfnamefont{J.~G.} \bibnamefont{Korner}},
  \bibnamefont{and} \bibinfo{author}{\bibfnamefont{O.~I.}
  \bibnamefont{Yakovlev}}, \bibinfo{journal}{Phys. Rev.}
  \textbf{\bibinfo{volume}{D55}}, \bibinfo{pages}{3016} (\bibinfo{year}{1997}),
  \eprint{hep-ph/9609469}.

\bibitem[{\citenamefont{Zhang and Huang}(2008)}]{Zhang:2008pm}
\bibinfo{author}{\bibfnamefont{J.-R.} \bibnamefont{Zhang}} \bibnamefont{and}
  \bibinfo{author}{\bibfnamefont{M.-Q.} \bibnamefont{Huang}},
  \bibinfo{journal}{Phys.Rev.} \textbf{\bibinfo{volume}{D78}},
  \bibinfo{pages}{094015} (\bibinfo{year}{2008}), \eprint{0811.3266}.

\bibitem[{\citenamefont{Flynn et~al.}(2003)\citenamefont{Flynn, Mescia, and
  Tariq}}]{Flynn:2003vz}
\bibinfo{author}{\bibfnamefont{J.}~\bibnamefont{Flynn}},
  \bibinfo{author}{\bibfnamefont{F.}~\bibnamefont{Mescia}}, \bibnamefont{and}
  \bibinfo{author}{\bibfnamefont{A.~S.~B.} \bibnamefont{Tariq}}
  (\bibinfo{collaboration}{UKQCD Collaboration}), \bibinfo{journal}{JHEP}
  \textbf{\bibinfo{volume}{0307}}, \bibinfo{pages}{066} (\bibinfo{year}{2003}),
  \eprint{hep-lat/0307025}.

\bibitem[{\citenamefont{Mathur et~al.}(2002)\citenamefont{Mathur, Lewis, and
  Woloshyn}}]{Mathur:2002ce}
\bibinfo{author}{\bibfnamefont{N.}~\bibnamefont{Mathur}},
  \bibinfo{author}{\bibfnamefont{R.}~\bibnamefont{Lewis}}, \bibnamefont{and}
  \bibinfo{author}{\bibfnamefont{R.}~\bibnamefont{Woloshyn}},
  \bibinfo{journal}{Phys.Rev.} \textbf{\bibinfo{volume}{D66}},
  \bibinfo{pages}{014502} (\bibinfo{year}{2002}), \eprint{hep-ph/0203253}.

\bibitem[{\citenamefont{Lewis et~al.}(2001)\citenamefont{Lewis, Mathur, and
  Woloshyn}}]{Lewis:2001iz}
\bibinfo{author}{\bibfnamefont{R.}~\bibnamefont{Lewis}},
  \bibinfo{author}{\bibfnamefont{N.}~\bibnamefont{Mathur}}, \bibnamefont{and}
  \bibinfo{author}{\bibfnamefont{R.}~\bibnamefont{Woloshyn}},
  \bibinfo{journal}{Phys.Rev.} \textbf{\bibinfo{volume}{D64}},
  \bibinfo{pages}{094509} (\bibinfo{year}{2001}), \eprint{hep-ph/0107037}.

\bibitem[{\citenamefont{Liu et~al.}(2010)\citenamefont{Liu, Lin, Orginos, and
  Walker-Loud}}]{Liu:2009jc}
\bibinfo{author}{\bibfnamefont{L.}~\bibnamefont{Liu}},
  \bibinfo{author}{\bibfnamefont{H.-W.} \bibnamefont{Lin}},
  \bibinfo{author}{\bibfnamefont{K.}~\bibnamefont{Orginos}}, \bibnamefont{and}
  \bibinfo{author}{\bibfnamefont{A.}~\bibnamefont{Walker-Loud}},
  \bibinfo{journal}{Phys. Rev.} \textbf{\bibinfo{volume}{D81}},
  \bibinfo{pages}{094505} (\bibinfo{year}{2010}), \eprint{0909.3294}.

\bibitem[{\citenamefont{Namekawa et~al.}(2011)}]{Namekawa:2011wt}
\bibinfo{author}{\bibfnamefont{Y.}~\bibnamefont{Namekawa}} \bibnamefont{et~al.}
  (\bibinfo{collaboration}{PACS-CS Collaboration}),
  \bibinfo{journal}{Phys.Rev.} \textbf{\bibinfo{volume}{D84}},
  \bibinfo{pages}{074505} (\bibinfo{year}{2011}), \eprint{1104.4600}.

\bibitem[{\citenamefont{Alexandrou
  et~al.}(2012{\natexlab{a}})\citenamefont{Alexandrou, Carbonell, Christaras,
  Drach, Gravina et~al.}}]{Alexandrou:2012xk}
\bibinfo{author}{\bibfnamefont{C.}~\bibnamefont{Alexandrou}},
  \bibinfo{author}{\bibfnamefont{J.}~\bibnamefont{Carbonell}},
  \bibinfo{author}{\bibfnamefont{D.}~\bibnamefont{Christaras}},
  \bibinfo{author}{\bibfnamefont{V.}~\bibnamefont{Drach}},
  \bibinfo{author}{\bibfnamefont{M.}~\bibnamefont{Gravina}},
  \bibnamefont{et~al.}, \bibinfo{journal}{Phys.Rev.}
  \textbf{\bibinfo{volume}{D86}}, \bibinfo{pages}{114501}
  (\bibinfo{year}{2012}{\natexlab{a}}), \eprint{1205.6856}.

\bibitem[{\citenamefont{Briceno et~al.}(2012)\citenamefont{Briceno, Lin, and
  Bolton}}]{Briceno:2012wt}
\bibinfo{author}{\bibfnamefont{R.~A.} \bibnamefont{Briceno}},
  \bibinfo{author}{\bibfnamefont{H.-W.} \bibnamefont{Lin}}, \bibnamefont{and}
  \bibinfo{author}{\bibfnamefont{D.~R.} \bibnamefont{Bolton}},
  \bibinfo{journal}{Phys.Rev.} \textbf{\bibinfo{volume}{D86}},
  \bibinfo{pages}{094504} (\bibinfo{year}{2012}), \eprint{1207.3536}.

\bibitem[{\citenamefont{Namekawa et~al.}(2013)}]{Namekawa:2013vu}
\bibinfo{author}{\bibfnamefont{Y.}~\bibnamefont{Namekawa}} \bibnamefont{et~al.}
  (\bibinfo{collaboration}{PACS-CS Collaboration}) (\bibinfo{year}{2013}),
  \eprint{1301.4743}.

\bibitem[{\citenamefont{Collins et~al.}(2011)\citenamefont{Collins, Gockeler,
  Hagler, Horsley, Nakamura et~al.}}]{Collins:2011mk}
\bibinfo{author}{\bibfnamefont{S.}~\bibnamefont{Collins}},
  \bibinfo{author}{\bibfnamefont{M.}~\bibnamefont{Gockeler}},
  \bibinfo{author}{\bibfnamefont{P.}~\bibnamefont{Hagler}},
  \bibinfo{author}{\bibfnamefont{R.}~\bibnamefont{Horsley}},
  \bibinfo{author}{\bibfnamefont{Y.}~\bibnamefont{Nakamura}},
  \bibnamefont{et~al.}, \bibinfo{journal}{Phys.Rev.}
  \textbf{\bibinfo{volume}{D84}}, \bibinfo{pages}{074507}
  (\bibinfo{year}{2011}), \eprint{1106.3580}.

\bibitem[{\citenamefont{Alexandrou et~al.}(2011)\citenamefont{Alexandrou,
  Brinet, Carbonell, Constantinou, Harraud et~al.}}]{Alexandrou:2011db}
\bibinfo{author}{\bibfnamefont{C.}~\bibnamefont{Alexandrou}},
  \bibinfo{author}{\bibfnamefont{M.}~\bibnamefont{Brinet}},
  \bibinfo{author}{\bibfnamefont{J.}~\bibnamefont{Carbonell}},
  \bibinfo{author}{\bibfnamefont{M.}~\bibnamefont{Constantinou}},
  \bibinfo{author}{\bibfnamefont{P.}~\bibnamefont{Harraud}},
  \bibnamefont{et~al.}, \bibinfo{journal}{Phys.Rev.}
  \textbf{\bibinfo{volume}{D83}}, \bibinfo{pages}{094502}
  (\bibinfo{year}{2011}), \eprint{1102.2208}.

\bibitem[{\citenamefont{Liu and Oka}(2012)}]{Liu:2011xc}
\bibinfo{author}{\bibfnamefont{Y.-R.} \bibnamefont{Liu}} \bibnamefont{and}
  \bibinfo{author}{\bibfnamefont{M.}~\bibnamefont{Oka}},
  \bibinfo{journal}{Phys.Rev.} \textbf{\bibinfo{volume}{D85}},
  \bibinfo{pages}{014015} (\bibinfo{year}{2012}), \eprint{1103.4624}.

\bibitem[{\citenamefont{Meguro et~al.}(2011)\citenamefont{Meguro, Liu, and
  Oka}}]{Meguro:2011nr}
\bibinfo{author}{\bibfnamefont{W.}~\bibnamefont{Meguro}},
  \bibinfo{author}{\bibfnamefont{Y.-R.} \bibnamefont{Liu}}, \bibnamefont{and}
  \bibinfo{author}{\bibfnamefont{M.}~\bibnamefont{Oka}},
  \bibinfo{journal}{Phys.Lett.} \textbf{\bibinfo{volume}{B704}},
  \bibinfo{pages}{547} (\bibinfo{year}{2011}), \eprint{1105.3693}.

\bibitem[{\citenamefont{Julia-Diaz and Riska}(2004)}]{JuliaDiaz:2004vh}
\bibinfo{author}{\bibfnamefont{B.}~\bibnamefont{Julia-Diaz}} \bibnamefont{and}
  \bibinfo{author}{\bibfnamefont{D.}~\bibnamefont{Riska}},
  \bibinfo{journal}{Nucl.Phys.} \textbf{\bibinfo{volume}{A739}},
  \bibinfo{pages}{69} (\bibinfo{year}{2004}), \eprint{hep-ph/0401096}.

\bibitem[{\citenamefont{Faessler et~al.}(2006)\citenamefont{Faessler, Gutsche,
  Ivanov, Korner, Lyubovitskij et~al.}}]{Faessler:2006ft}
\bibinfo{author}{\bibfnamefont{A.}~\bibnamefont{Faessler}},
  \bibinfo{author}{\bibfnamefont{T.}~\bibnamefont{Gutsche}},
  \bibinfo{author}{\bibfnamefont{M.}~\bibnamefont{Ivanov}},
  \bibinfo{author}{\bibfnamefont{J.}~\bibnamefont{Korner}},
  \bibinfo{author}{\bibfnamefont{V.}~\bibnamefont{Lyubovitskij}},
  \bibnamefont{et~al.}, \bibinfo{journal}{Phys.Rev.}
  \textbf{\bibinfo{volume}{D73}}, \bibinfo{pages}{094013}
  (\bibinfo{year}{2006}), \eprint{hep-ph/0602193}.

\bibitem[{\citenamefont{Albertus et~al.}(2007)\citenamefont{Albertus,
  Hernandez, Nieves, and Verde-Velasco}}]{Albertus:2006ya}
\bibinfo{author}{\bibfnamefont{C.}~\bibnamefont{Albertus}},
  \bibinfo{author}{\bibfnamefont{E.}~\bibnamefont{Hernandez}},
  \bibinfo{author}{\bibfnamefont{J.}~\bibnamefont{Nieves}}, \bibnamefont{and}
  \bibinfo{author}{\bibfnamefont{J.}~\bibnamefont{Verde-Velasco}},
  \bibinfo{journal}{Eur.Phys.J.} \textbf{\bibinfo{volume}{A32}},
  \bibinfo{pages}{183} (\bibinfo{year}{2007}), \eprint{hep-ph/0610030}.

\bibitem[{\citenamefont{Zhu et~al.}(1997)\citenamefont{Zhu, Hwang, and
  Yang}}]{Zhu:1997as}
\bibinfo{author}{\bibfnamefont{S.-L.} \bibnamefont{Zhu}},
  \bibinfo{author}{\bibfnamefont{W.-Y.} \bibnamefont{Hwang}}, \bibnamefont{and}
  \bibinfo{author}{\bibfnamefont{Z.-S.} \bibnamefont{Yang}},
  \bibinfo{journal}{Phys.Rev.} \textbf{\bibinfo{volume}{D56}},
  \bibinfo{pages}{7273} (\bibinfo{year}{1997}), \eprint{hep-ph/9708411}.

\bibitem[{\citenamefont{Aliev et~al.}(2002)\citenamefont{Aliev, Ozpineci, and
  Savci}}]{Aliev:2001ig}
\bibinfo{author}{\bibfnamefont{T.}~\bibnamefont{Aliev}},
  \bibinfo{author}{\bibfnamefont{A.}~\bibnamefont{Ozpineci}}, \bibnamefont{and}
  \bibinfo{author}{\bibfnamefont{M.}~\bibnamefont{Savci}},
  \bibinfo{journal}{Phys.Rev.} \textbf{\bibinfo{volume}{D65}},
  \bibinfo{pages}{056008} (\bibinfo{year}{2002}), \eprint{hep-ph/0107196}.

\bibitem[{\citenamefont{Can et~al.}(2013)\citenamefont{Can, Erkol, Oka,
  Ozpineci, and Takahashi}}]{Can:2012tx}
\bibinfo{author}{\bibfnamefont{K.}~\bibnamefont{Can}},
  \bibinfo{author}{\bibfnamefont{G.}~\bibnamefont{Erkol}},
  \bibinfo{author}{\bibfnamefont{M.}~\bibnamefont{Oka}},
  \bibinfo{author}{\bibfnamefont{A.}~\bibnamefont{Ozpineci}}, \bibnamefont{and}
  \bibinfo{author}{\bibfnamefont{T.}~\bibnamefont{Takahashi}},
  \bibinfo{journal}{Phys.Lett.} \textbf{\bibinfo{volume}{B719}},
  \bibinfo{pages}{103} (\bibinfo{year}{2013}), \eprint{1210.0869}.

\bibitem[{\citenamefont{Aoki et~al.}(2009)}]{Aoki:2008sm}
\bibinfo{author}{\bibfnamefont{S.}~\bibnamefont{Aoki}} \bibnamefont{et~al.}
  (\bibinfo{collaboration}{PACS-CS}), \bibinfo{journal}{Phys. Rev.}
  \textbf{\bibinfo{volume}{D79}}, \bibinfo{pages}{034503}
  (\bibinfo{year}{2009}), \eprint{0807.1661}.

\bibitem[{\citenamefont{Bali et~al.}(2011)\citenamefont{Bali, Collins, and
  Ehmann}}]{Bali:2011rd}
\bibinfo{author}{\bibfnamefont{G.~S.} \bibnamefont{Bali}},
  \bibinfo{author}{\bibfnamefont{S.}~\bibnamefont{Collins}}, \bibnamefont{and}
  \bibinfo{author}{\bibfnamefont{C.}~\bibnamefont{Ehmann}},
  \bibinfo{journal}{Phys.Rev.} \textbf{\bibinfo{volume}{D84}},
  \bibinfo{pages}{094506} (\bibinfo{year}{2011}), \eprint{1110.2381}.

\bibitem[{\citenamefont{Brodsky et~al.}(2011)\citenamefont{Brodsky, Guo, Hanhart, and
  Mei{\ss}ner}}]{Brodsky:2011zs}
\bibinfo{author}{\bibfnamefont{S.~J.} \bibnamefont{Brodsky}},
  \bibinfo{author}{\bibfnamefont{F.-K.}~\bibnamefont{Guo}},
  \bibinfo{author}{\bibfnamefont{C.}~\bibnamefont{Hanhart}}, \bibnamefont{and}
  \bibinfo{author}{\bibfnamefont{U.-G.}~\bibnamefont{Mei{\ss}ner}},
  \bibinfo{journal}{Phys.Lett.} \textbf{\bibinfo{volume}{B698}},
  \bibinfo{pages}{251-255} (\bibinfo{year}{2011}), \eprint{hep-ph/1101.1983}.

\bibitem[{\citenamefont{El-Khadra et~al.}(1997)\citenamefont{El-Khadra,
  Kronfeld, and Mackenzie}}]{ElKhadra:1996mp}
\bibinfo{author}{\bibfnamefont{A.~X.} \bibnamefont{El-Khadra}},
  \bibinfo{author}{\bibfnamefont{A.~S.} \bibnamefont{Kronfeld}},
  \bibnamefont{and} \bibinfo{author}{\bibfnamefont{P.~B.}
  \bibnamefont{Mackenzie}}, \bibinfo{journal}{Phys. Rev.}
  \textbf{\bibinfo{volume}{D55}}, \bibinfo{pages}{3933} (\bibinfo{year}{1997}),
  \eprint{hep-lat/9604004}.

\bibitem[{\citenamefont{Burch et~al.}(2010)\citenamefont{Burch, DeTar,
  Di~Pierro, El-Khadra, Freeland et~al.}}]{Burch:2009az}
\bibinfo{author}{\bibfnamefont{T.}~\bibnamefont{Burch}},
  \bibinfo{author}{\bibfnamefont{C.}~\bibnamefont{DeTar}},
  \bibinfo{author}{\bibfnamefont{M.}~\bibnamefont{Di~Pierro}},
  \bibinfo{author}{\bibfnamefont{A.}~\bibnamefont{El-Khadra}},
  \bibinfo{author}{\bibfnamefont{E.}~\bibnamefont{Freeland}},
  \bibnamefont{et~al.}, \bibinfo{journal}{Phys.Rev.}
  \textbf{\bibinfo{volume}{D81}}, \bibinfo{pages}{034508}
  (\bibinfo{year}{2010}), \eprint{0912.2701}.

\bibitem[{\citenamefont{Alexandrou
  et~al.}(2012{\natexlab{b}})\citenamefont{Alexandrou, Hadjiyiannakou, Koutsou,
  O'Cais, and Strelchenko}}]{Alexandrou:2012zz}
\bibinfo{author}{\bibfnamefont{C.}~\bibnamefont{Alexandrou}},
  \bibinfo{author}{\bibfnamefont{K.}~\bibnamefont{Hadjiyiannakou}},
  \bibinfo{author}{\bibfnamefont{G.}~\bibnamefont{Koutsou}},
  \bibinfo{author}{\bibfnamefont{A.}~\bibnamefont{O'Cais}}, \bibnamefont{and}
  \bibinfo{author}{\bibfnamefont{A.}~\bibnamefont{Strelchenko}},
  \bibinfo{journal}{Comput.Phys.Commun.} \textbf{\bibinfo{volume}{183}},
  \bibinfo{pages}{1215} (\bibinfo{year}{2012}{\natexlab{b}}),
  \eprint{1108.2473}.

\bibitem[{\citenamefont{Beringer et~al.}(2012)\citenamefont{Beringer, Arguin,
  Barnett, Copic, Dahl, Groom, Lin, Lys, Murayama, Wohl
  et~al.}}]{PhysRevD.86.010001}
\bibinfo{author}{\bibfnamefont{J.}~\bibnamefont{Beringer}},
  \bibinfo{author}{\bibfnamefont{J.~F.} \bibnamefont{Arguin}},
  \bibinfo{author}{\bibfnamefont{R.~M.} \bibnamefont{Barnett}},
  \bibinfo{author}{\bibfnamefont{K.}~\bibnamefont{Copic}},
  \bibinfo{author}{\bibfnamefont{O.}~\bibnamefont{Dahl}},
  \bibinfo{author}{\bibfnamefont{D.~E.} \bibnamefont{Groom}},
  \bibinfo{author}{\bibfnamefont{C.~J.} \bibnamefont{Lin}},
  \bibinfo{author}{\bibfnamefont{J.}~\bibnamefont{Lys}},
  \bibinfo{author}{\bibfnamefont{H.}~\bibnamefont{Murayama}},
  \bibinfo{author}{\bibfnamefont{C.~G.} \bibnamefont{Wohl}},
  \bibnamefont{et~al.} (\bibinfo{collaboration}{Particle Data Group}),
  \bibinfo{journal}{Phys. Rev. D} \textbf{\bibinfo{volume}{86}},
  \bibinfo{pages}{010001} (\bibinfo{year}{2012}).

\bibitem[{\citenamefont{Yamamoto and Suganuma}(2008)}]{Yamamoto:2007nn}
\bibinfo{author}{\bibfnamefont{A.}~\bibnamefont{Yamamoto}} \bibnamefont{and}
  \bibinfo{author}{\bibfnamefont{H.}~\bibnamefont{Suganuma}},
  \bibinfo{journal}{Phys.Rev.} \textbf{\bibinfo{volume}{D77}},
  \bibinfo{pages}{014036} (\bibinfo{year}{2008}), \eprint{0709.0171}.

\bibitem[{\citenamefont{Sakurai}(1969)}]{Sakurai:69}
\bibinfo{author}{\bibfnamefont{J.~J.} \bibnamefont{Sakurai}},
  \emph{\bibinfo{title}{Currents and mesons}} (\bibinfo{publisher}{University
  of Chicago Press}, \bibinfo{address}{Chicago}, \bibinfo{year}{1969}).

\bibitem[{\citenamefont{Edwards and Joo}(2005)}]{Edwards:2004sx}
\bibinfo{author}{\bibfnamefont{R.~G.} \bibnamefont{Edwards}} \bibnamefont{and}
  \bibinfo{author}{\bibfnamefont{B.}~\bibnamefont{Joo}}
  (\bibinfo{collaboration}{SciDAC Collaboration, LHPC Collaboration, UKQCD
  Collaboration}), \bibinfo{journal}{Nucl.Phys.Proc.Suppl.}
  \textbf{\bibinfo{volume}{140}}, \bibinfo{pages}{832} (\bibinfo{year}{2005}),
  \eprint{hep-lat/0409003}.

\end{thebibliography}
\end{document}